\documentclass[aps,prd,preprint,nofootinbib]{revtex4}
\usepackage{amsfonts,amsmath,amssymb}
\usepackage{mathrsfs}
\usepackage{graphicx}
\usepackage{multirow}
\usepackage{epsfig}
\usepackage{graphicx}
\usepackage{subfigure}
\usepackage{booktabs}
\usepackage{array}
\usepackage{tabularx}
\usepackage{slashed}
\usepackage{braket}
\usepackage{bm}
\usepackage{diagbox}
\usepackage{array}
\usepackage{makecell}

\begin{document}
\baselineskip 20pt
\title{Charmonium pair production in ultraperipheral collision}
\author{\vspace{1cm} Hao Yang$^1$\footnote[1]{yanghao2023@scu.edu.cn}, Zi-Qiang Chen$^{2}$\footnote[2]{chenziqiang13@gzhu.edu.cn, corresponding author} and Bingwei Long$^{1,3}$\footnote[3]{bingwei@scu.edu.cn, corresponding author}\\}

\affiliation{
$^1$College of Physics, Sichuan University, Chengdu, Sichuan 610065, China\\
$^2$School of Physics and Materials Science, Guangzhou University, Guangzhou 510006, China\\
$^3$Southern Center for Nuclear-Science Theory (SCNT), Institute of Modern Physics, Chinese Academy of Sciences, Huizhou 516000, Guangdong, China\vspace{0.6cm}
}

\begin{abstract}
We study the exclusive double charmonium ($J/\psi \mbox{-} J/\psi$ and $\eta_c \mbox{-} \eta_c$) production through photon-photon fusion via ultraperipheral collision (UPC) at the HL-LHC and FCC with next-to-leading order (NLO) QCD predictions in the framework of non-relativistic QCD (NRQCD). Numerical results indicate that the NLO corrections for $J/\psi$ pair are large and negative, while positive for $\eta_c$ pair. The total cross section of $J/\psi \mbox{-} J/\psi$ ($\eta_c \mbox{-} \eta_c$) in Pb-Pb UPC is 28.0 (65.1) nb at nucleon-nucleon c.m. energy $\sqrt{s_{NN}} = 5.52$ TeV. Due to the backgrounds from various QCD interactions at UPC are highly suppressed and the event topologies for charmonium pair are easy to tag, the phenomenological studies at the LHC and FCC are feasible. The detailed transverse momentum $p_T$, diphoton invariant mass $m_{\gamma\gamma}$ and the rapidity difference $\Delta y$ distributions are given. The production for X(6900) is also discussed.     

\end{abstract}

\maketitle

\section{INTRODUCTION}
As the simplest hadron consisting of two heavy quarks, heavy quarkonium provides an ideal laboratory to probe the perturbative and nonperturbative properties of quantum chromodynamics (QCD).
Due to the large mass of the constituent quarks, quarkonium production process is assumed to be factorizable into two stages. First, a heavy quark-antiquark pair whose invariant mass near the bound-state mass is produced perturbatively, then the pair binds into a quarkonium state nonperturbatively. The non-relativistic QCD (NRQCD) factorization formalism, which is proposed by Bodwin, Braaten, and Lepage \cite{Bodwin:1994jh}, provides a systematic framework for the theoretical study of quarkonium production. In NRQCD factorization approach, the nonperturbative hadronization effects are encoded into the long distance matrix elements (LDMEs), which are sorted by means of velocity scaling rules. As a result, the theoretical prediction takes the form of a double expansion in the strong coupling constant $\alpha_s$ and the heavy quark relative velocity $v$.

Quarkonium production in photon-photon collisions is a compelling area of study, offering a sensitive probe for testing NRQCD factorization formalism. The cross section for inclusive $J/\psi$ production via photon-photon scattering was measured by the LEP II experiment \cite{DELPHI:2003hen}, yet the underlying production mechanisms remain incompletely understood. Theoretical researches indicate that the color-singlet (CS) sectors are inadequate to explain the experimental data, while the color-octet (CO) contributions based on different LDME sets diverse quite a lot, which even lead to opposite conclusions in some cases \cite{Ma:1997bi,Japaridze:1998ss,Klasen:2001mi,Klasen:2001cu,Qiao:2003ba,Li:2009zzu,Chen:2016hju,Butenschoen:2011yh}. On the other hand, with the advent of high-luminosity colliders, some exclusive processes, like the $J/\psi$-pair production in photon-photon collision, can be measured with high accuracy. In comparison with $J/\psi$ inclusive production, the production mechanism of $J/\psi$-pair is much simpler, due to the fact that the CO contributions, which are suppressed by $v^8$, are expected to be insignificant. Therefore, more clear conclusions can be expected from the study of this exclusive process.

In our previous work \cite{Yang:2020xkl}, we calculated the next-to-leading order (NLO) QCD corrections to the $\gamma+\gamma\to J/\psi+J/\psi$ process, and investigated the corresponding cross section at the SuperKEKB $e^+e^-$ collider. It should be remarked that the ultraperipheral collision (UPC) of heavy ions is an important alternative test window to study the photon-photon physics.
In the UPC, the ion impact parameter is larger than twice the ion radius, hence the ions are kept unbroken during the collision, and the interaction is pure electromagnetic. Due to the coherent action of all the protons in the nucleus, the electromagnetic field is very strong, and the resulting flux of equivalent photons is large. Explicitly, the photon flux from the nuclei scaling as the square of the charge number $Z^2$, giving a two-photon luminosity scaling as $Z^4$, which eventually leads to a large cross section. In addition, compared with photon-photon collision from electron bremsstrahlung, the UPC event topologies are featured as lacking pileup and large rapidity gap, hence the event signals can be reconstructed with high efficiency. For these reasons, in this work, we investigate the $\gamma+\gamma\to J/\psi+J/\psi$ process at the heavy ion UPC.

The rest of this paper is organized as follows. 
In Sect. II, we present the primary formulas employed in the calculation. 
In Sect. III,  the total cross sections, their uncertainties, and the differential distributions are discussed. The last section is reserved for summary and conclusions.

\section{FORMULATION}
\begin{figure}[htbp!]			
	\centering
	\subfigure{\includegraphics[scale=0.7]{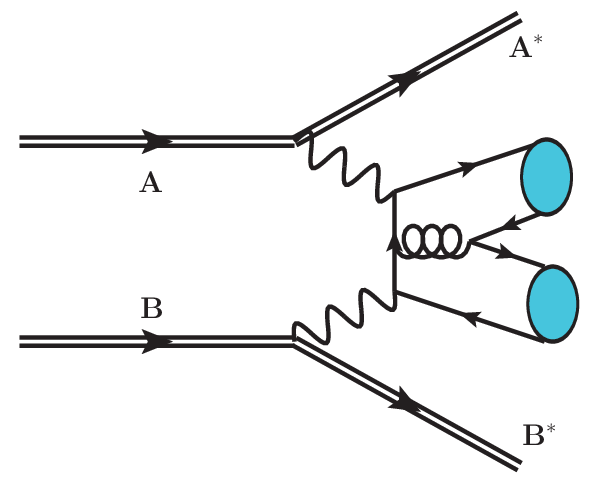}}
	\caption{The schematic diagram for quarkonium pair production via photon-photon fusion in an UPC of ions.}
	\label{Figfeyn}
\end{figure}

The process of quarkonium pair production via photon-photon fusion in an UPC of ions A and B is displayed in FIG. \ref{Figfeyn}.
In the framework of equivalent photon approximation (EPA) \cite{vonWeizsacker:1934nji,Williams:1934ad}, the total cross section can be expressed as the convolution of the $\gamma+\gamma\to \mathcal{H}+\mathcal{H}$ cross section with the equivalent photon spectral functions
\begin{align}
	\sigma(A+B\to A+B+\mathcal{H}+\mathcal{H}) &= \int \dfrac{dx_1}{x_1} \dfrac{dx_2}{x_2}n_1(x_1)n_2(x_2) \hat{\sigma}(\gamma+\gamma\to \mathcal{H}+\mathcal{H}),
	\label{eq_fm}
\end{align}
where $x_i=E_{\gamma,i}/E_\text{beam}$ is the energy fraction of each photon, $n_i(x)$ is the photon spectral function.
For ion with charge $Z$, the photon spectral function takes the form \cite{Cahn:1990jk}
\begin{align}
n(x) = \dfrac{2 Z^2 \alpha}{\pi}\left\{\chi K_0(\chi)K_1(\chi) -(1-\gamma_\text{L}^{-2}) \dfrac{\chi^2}{2}\left[K_1^2(\chi)-K_0^2(\chi)\right]\right\},
\end{align}
where $K_i$ is the modified Bessel function, $\gamma_\text{L}=E_\text{beam}/m_\text{p}$ is the Lorentz boost factor with $m_\text{p}$ denotes the proton mass.
The variable $x$ is absorbed in to $\chi=xm_\text{p}b_\text{min}$, in which the minimum of the impact parameter $b_\text{min}$ will be set as the nucleus radius.

In our previous work \cite{Yang:2020xkl}, we calculated the NLO QCD corrections to the $\gamma+\gamma\to \mathcal{H}+\mathcal{H}$ process.
In the following, we briefly review the ingredients of the calculation.

Up to NLO, the $\hat{\sigma}(\gamma + \gamma \rightarrow \mathcal{H}+\mathcal{H})$ of Eq. (\ref{eq_fm})  incorporate the LO contribution and the NLO correction term, that is
\begin{equation}
	\hat{\sigma}( \gamma + \gamma \rightarrow \mathcal{H}+\mathcal{H}) = \hat{\sigma}_{\rm born} + \hat{\sigma}_{\rm NLO}+\mathcal{O}(\alpha^{2}\alpha^{4}_{s})\ ,
\end{equation}
where
\begin{equation}
	\begin{split}
		&\hat{\sigma}_{\rm born}=\frac{1}{2}\frac{1}{2\hat{s}}\int\overline{\sum}|\mathcal{M}_{\rm tree}|^{2}d{\rm PS}_{2}\ ,\\
		&\hat{\sigma}_{\rm NLO}=\frac{1}{2}\frac{1}{2\hat{s}}\int\overline{\sum}2{\rm Re}(\mathcal{M}^{*}_{\rm tree}\mathcal{M}_{\rm oneloop})d{\rm PS}_{2}\ .
	\end{split}
\end{equation}
Here, $\frac{1}{2}$ is the statistical factor for identical particles in the final state, $\hat{s}$ is the center-of-mass energy square of the two-photon system, $\overline{\sum}$ means sum (average) over the polarizations and colors of final (initial) state, $d{\rm PS}_{2}$ denotes two-body phase space. 

In the calculation of the amplitude, the covariant projection method is applied.
The standard form of the spin and color projector is \cite{Bodwin:2002cfe}:
\begin{equation}
	v(p_{\bar{c}})\bar{u}(p_c)\to \frac{1}{4\sqrt{2}E(E+m_c)}(\not\! p_{\bar{c}}-m_c)\not\! \epsilon^{*}_{S}(\not\!P+2E)(\not\!p_c+m_c)\otimes(\frac{\bf{1}_{c}}{\sqrt{N_{c}}})\,
\end{equation}
where $\epsilon^{*}_{S}$ denotes the polarization vector of the spin-triplet state, $p_c$ and $p_{\bar{c}}$ are the momenta of heavy quark and antiquark respectively, $P=p_c+p_{\bar{c}}$ is the momentum of quarkonium, $E=\sqrt{P^{2}/4}$ is the energy of heavy (anti) quark in quarkonium rest frame, $\bf{1}_{c}$ represents the unit color matrix, and $N_c=3$ is the colors number in QCD. 
At the leading order of the relative velocity expansion, it is legitimate to take the approximation of $p_c=p_{\bar{c}}=P/2$ and $E=m_c$. 
As for the spin-singlet state, the projection operator can be obtained by replacing the $\not\! \epsilon^{*}_{S}$ with $\gamma_5$.

In the calculation of the one-loop amplitude, we encounter ultraviolet (UV) and infrared (IR) singularities.
Both of them are regularized by the dimensional regularization with $D=4-2\epsilon$.
According to Ref.\cite{Bodwin:1994jh}, the soft IR singularities are canceled with each other, and the collinear IR singularity in the hard-collinear region will not emerge due to the final charm quarks are massive. 
The Coulomb IR singularities, where a potential gluon is exchanged between the constituent quarks of a quarkonium, vanish in the dimensional regularization since we set $p^{}_Q=p^{}_{\bar{Q}}$ before calculating the Feynman integrals \cite{Beneke:1997zp}. 
As for the UV singularities, they are removed by the standard renormalization procedure.
The relevant renormalization constants include $Z_2$, $Z_m$, and $Z_g$, corresponding to heavy quark field, heavy quark mass, and strong coupling constant, respectively.
We define $Z_2$, $Z_m$ in the on-mass-shell (OS) scheme, $Z_g$ in the modified minimal-subtraction ($\overline{\rm MS}$) scheme.
The counter terms are
\begin{equation}
	\begin{split}
		&\delta Z^{\rm OS}_{2}=-C_{F}\frac{\alpha_{s}}{4\pi}\left[\frac{1}{\epsilon_{\rm UV}}+\frac{2}{\epsilon_{\rm IR}}-3\gamma_{E}+3\ln\frac{4\pi\mu^{2}}{m_Q^{2}}+4\right],\\
		&\delta Z^{\rm OS}_{m}=-3C_{F}\frac{\alpha_{s}}{4\pi}\left[\frac{1}{\epsilon_{\rm UV}}-\gamma_{E}+\ln\frac{4\pi\mu^{2}}{m_Q^{2}}+\frac{4}{3}\right],\\
		&\delta Z^{\overline{\rm MS}}_{g}=-\frac{\beta_{0}}{2}\frac{\alpha_{s}}{4\pi}\left[\frac{1}{\epsilon_{\rm UV}}-\gamma_{E}+\ln(4\pi)\right].
		\label{eq_ctterm}
	\end{split}
\end{equation}
Here, $\mu$ is the renormalization scale, $\gamma_{E}$ is the Euler's constant, $\beta_{0}=\frac{11}{3}C_{A}-\frac{4}{3}T_{F}n_{f}$ is the one-loop coefficient of the QCD $\beta$-function, and $n_{f}$ denotes the active quark flavor numbers. 
The color factors $C_{A}=3, C_{F}=\frac{4}{3}$ and $T_{F}=\frac{1}{2}$.

\section{NUMERICAL RESULTS}
The input parameters taken in the numerical calculation go as follows:
\begin{align}
&\alpha=1/137.065,\ m_c=1.5\pm 0.1\ {\rm GeV},\ m_\text{p}=0.9315\ {\rm GeV},\nonumber \\
&|R_{J/\psi}^{\rm LO}(0)|^2=0.528\ {\rm GeV}^3,\ |R_{J/\psi}^{\rm NLO}(0)|^2=0.907\ {\rm GeV}^3,\ R_{\eta_c}(0)=R_{J/\psi}(0).
\end{align}
Here, the charm quark mass is set to be roughly one half of the $J/\psi$ or $\eta_c$ meson mass.
The $J/\psi$ radial wave function at the origin is extracted from its leptonic width
\begin{equation}
	\Gamma(J/\psi\to e^+e^-)=\frac{4\alpha^2}{9m_c^2}|R_H(0)|^2\left(1-4C_F\frac{\alpha_s(\mu_0)}{\pi}\right),
\end{equation}
with $\mu_0=2m^{}_c$, $\Gamma(J/\psi\to e^+e^-)=5.55$ keV \cite{ParticleDataGroup:2024cfk}.
According to the heavy quark spin symmetry of NRQCD at the leading order in relative velocity expansion \cite{Bodwin:1994jh}, here we take $R_{\eta_c}(0)=R_{J/\psi}(0)$.
Note that, in the LO calculation, LO extraction of the wave function and one-loop formula of the strong coupling are applied; while in the NLO, NLO extraction and two-loop formula of the strong coupling are adopted. 
The two-loop formula for the running coupling constant takes the form
\begin{equation}
	\frac{\alpha_{s}(\mu)}{4\pi}=\frac{1}{\beta_{0}L}-\frac{\beta_{1}\ln L}{\beta^{3}_{0}L^{2}},
	\label{eq_2cpl}
\end{equation}
where, $L=\ln (\mu^{2}/\Lambda^{2}_{\rm QCD})$, $\beta_0=\tfrac{11}{3}C_A-\tfrac{4}{3}T_Fn_f$, $\beta_{1}=\frac{34}{3}C^{2}_{A}-4C_{F}T_{F}n_{f}-\frac{20}{3}C_{A}T_{F}n_{f}$, with $n_f=4$, $\Lambda_{\rm QCD}=297$ MeV.
 The one-loop formula can be obtained by dropping the $\beta_1$ term of Eq. (\ref{eq_2cpl}).
 
We first consider the Pb-Pb collision at the HL-LHC, for which we set $\sqrt{s_\text{NN}}=5.52$ TeV, $R_\text{A}=7.1$ fm.
The corresponding LO and NLO cross sections at different renormalization scale are presented in Table \ref{TabScale}, wherein, the central values refer to the results at $m_c=1.5$ GeV, the superscripts and subscripts corresponding to the results at $m_c=1.4$ GeV and $m_c=1.6$ GeV, respectively.
It can be seen that, 
As indicated in the previous work, the NLO corrections are significant, and display different behaviors for $J/\psi$-pair and $\eta_c$-pair productions.
For $J/\psi$-pair production, the NLO corrections are negative at $\mu=2m_c$ and $\mu=\sqrt{4m_c^2+p_t^2}$, while positive at $\mu=\sqrt{\hat{s}_{\gamma\gamma}}$;

\begin{table}[ht]
	\caption{The total NLO (LO) cross sections (nb) for $J/\psi\mbox{-}J/\psi$ and $\eta_c\mbox{-}\eta_c$ versus different renormalization scale $\mu$ in ultraperipheral Pb-Pb collision at nucleon-nucleon c.m. energy $\sqrt{s_{NN}} = $ 5.52 TeV.}
	\begin{center}
		\centering
		\begin{tabular}{|m{2.5cm}<{\centering}|m{3.8cm}<{\centering}|m{3.8cm}<{\centering}|m{3.8cm}<{\centering}|}
			\toprule
			\hline
			$\mu$                  & $2m_c$      & $\sqrt{4m_c^2+p_t^2}$ & $\sqrt{\hat{s}_{\gamma\gamma}}$\\
			\hline
			$J/\psi\mbox{-}J/\psi$ & 11.0 (120)  & $28.0^{+18.4}_{-11.1}\; \left(111^{+109}_{-51.9}\right)$  & 83.8 (62.6)\\
			\hline
			$\eta_c\mbox{-}\eta_c$ & 66.6 (29.9) & $65.1^{+62.9}_{-30.6}\; \left(27.7^{+27.0}_{-13.0}\right)$  & 48.7 (16.0)\\
			\hline  		
		\end{tabular}
	\end{center}    
	\label{TabScale}
\end{table}
 
  \begin{table}[ht]
 	\caption{The nucleon-nucleon (NN) c.m. energy $\sqrt{s_{NN}}$, effective charge radius $R_A$ \cite{Shao:2022cly}, total LO (up) and NLO (down) cross sections for $J/\psi\mbox{-}J/\psi$, $\eta_c\mbox{-}\eta_c$ and integrated luminosity per typical run $\mathcal{L}_{int}$ for ultraperipheral collisions at HL-LHC and FCC.}
 	\begin{center}
 		\centering
 		\scalebox{0.8}{
 			\begin{tabular}{|c|c|c|c|c|c|c|c|c||c|c|c|}
 				\toprule
 				\hline
 				System                 & Pb-Pb & Xe-Xe & Kr-Kr & Ar-Ar & Ca-Ca & O-O & p-Pb     & p-p & Pb-Pb & p-Pb     & p-p\\
 				\hline
 				$\sqrt{s_{NN}}$\ (TeV) & 5.52  & 5.86  & 6.46  & 6.3   & 7.0   & 7.0 & 8.8      & 14  & 39.4  & 62.8     & 100\\
 				\hline
 				$R_A\ \rm (fm)$        & 7.1   & 6.1   & 5.1   & 4.1   & 4.1   & 3.1 & 0.7,\ 7.1 & 0.7 & 7.1   & 0.7,\ 7.1 & 0.7\\
 				\hline
 				$J/\psi\mbox{-}J/\psi$ & \makecell[c]{111 nb\\ 28.0 nb} & \makecell[c]{25.8 nb\\ 6.44 nb} & \makecell[c]{6.58 nb\\ 1.62 nb} & \makecell[c]{485 pb\\ 119 pb} & \makecell[c]{807 pb\\ 198 pb} & \makecell[c]{25.7 pb\\ 6.28 pb} & \makecell[c]{62.7 pb\\ 15.1 pb} & \makecell[c]{23.6 fb\\ 5.57 fb} & \makecell[c]{527 nb\\ 127 nb} & \makecell[c]{188 pb\\ 44.0 pb} & \makecell[c]{55.1 fb\\ 12.7 fb}\\
 				\hline
 				$\eta_c\mbox{-}\eta_c$ & \makecell[c]{27.7 nb\\ 65.1 nb} & \makecell[c]{6.40 nb\\ 15.0 nb} & \makecell[c]{1.62 nb\\ 3.80 nb} & \makecell[c]{119 pb\\ 280 pb} & \makecell[c]{197 pb\\ 464 pb} & \makecell[c]{6.28 pb\\ 14.7 pb} & \makecell[c]{15.2 pb\\ 35.6 pb} & \makecell[c]{5.66 fb\\ 13.2 fb} & \makecell[c]{127 nb\\ 298 nb} & \makecell[c]{44.8 pb\\ 104 pb} & \makecell[c]{13.0 fb\\ 30.5 fb}\\
 				\hline
 				$\mathcal{L}_{int}$    & $5\ \rm nb^{-1}$ & $30\ \rm nb^{-1}$ & $120\ \rm nb^{-1}$ & $1.1\ \rm pb^{-1}$ & $0.8\ \rm pb^{-1}$ & $12\ \rm pb^{-1}$ & $1\ \rm pb^{-1}$ & $150\ \rm fb^{-1}$ & $110\ \rm nb^{-1}$ & $29\ \rm pb^{-1}$ & $1\ \rm ab^{-1}$\\
 				\hline   		
 			\end{tabular}
 		}
 	\end{center}    
 	\label{TabUPC}
 \end{table}
The interaction of heavy ions at large impact parameters is purely electromagnetical, such an interaction can be treated as real photon-photon fusion where the flux is enhanced by $Z^2$, Z is the charge of ion. Compared with photon-photon collision from electron bremsstrahlung, the UPC event topologies are featured as lacking pileup and large rapidity gap, in this way the event signatures can be reconstructed with high efficiency. The generic characteristics of photon-photon fusion in ultraperipheral ion-ion collisions at HL-LHC \cite{Bruce:2018yzs,dEnterria:2022sut} and FCC \cite{Dainese:2016gch,FCC:2018vvp} energies are collected into TABLE. \ref{TabUPC}. 

The total LO and NLO cross sections for double $J/\psi$ and $\eta_c$ via various UPCs are given in TABLE.\ref{TabUPC}, here no any dynamical cut is imposed. As indicated in the previous work \cite{Qiao:2001wv,Yang:2020xkl,He:2024lrb}, the NLO QCD radiative and relativistic corrections both turn out to be significant and negative for $J/\psi$, whereas positive for $\eta_c$. The NLO K-factor is around 0.25  (2.35) for $J/\psi$ ($\eta_c$) through various ion-ion UPCs.

Supposing the integrated luminosities in TABLE. \ref{TabUPC}, the produced $J/\psi\mbox{-}J/\psi$ numbers via various heavy ion UPCs at the HL-LHC are 140-194, while the produced numbers can reach 835 for p-p collision due to its relative high luminosity. For $\eta_c\mbox{-}\eta_c$, the produced numbers increase to 325-456 for heavy ion case, and reach 1980 for p-p collision. In experiment, $J/\psi$ can be reconstructed  through its leptonic decays, and $\eta_c$ can be reconstructed through its $K\bar{K}\pi$ decay channel. According to Ref. \cite{ParticleDataGroup:2024cfk}, the branching ratios ${\rm Br}(J/\psi\to l^+l^-(l=e,\mu))=12\%$ and ${\rm Br}(\eta_c\to K\bar{K}\pi)=7.3\%$, and hence the numbers of reconstructed candidates per year would reach 2-12 for $J/\psi$-pair production and 1-10 for $\eta_c$-pair. Due to the event topologies for ultraperipheral collision are vary clear, the background from various QCD interactions can be suppressed, hence the experimental investigation are feasible. As the collision energies and luminosities are highly improved at the FCC, the yield events for $J/\psi$ pair will reach 180-200, leaving open opportunity for detailed study of differential distributions, and further provides new test window for NRQCD.
\begin{figure}[htbp!]			
	\centering
	\subfigure{\includegraphics[scale=0.28]{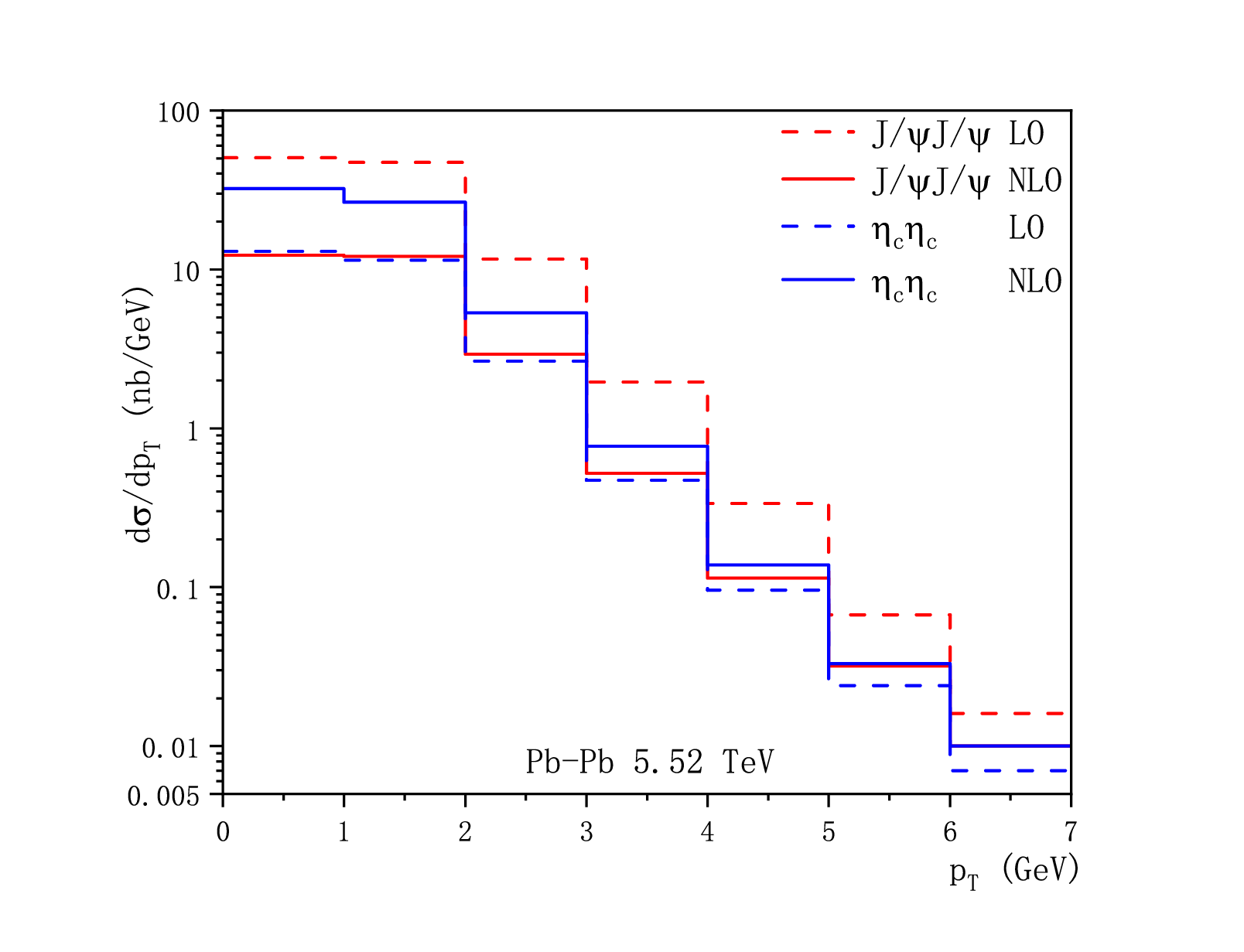}}
	\subfigure{\includegraphics[scale=0.28]{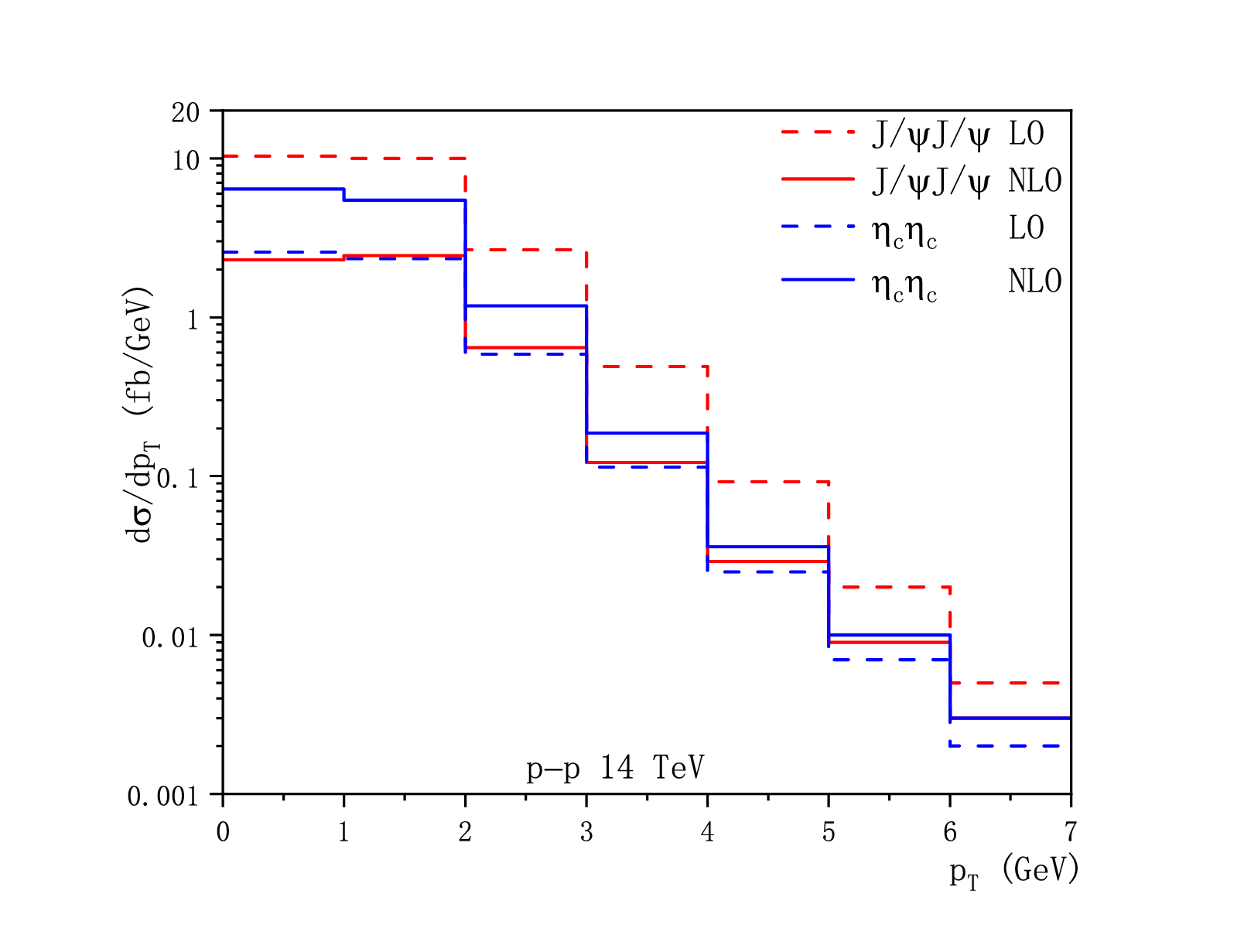}}
	\subfigure{\includegraphics[scale=0.28]{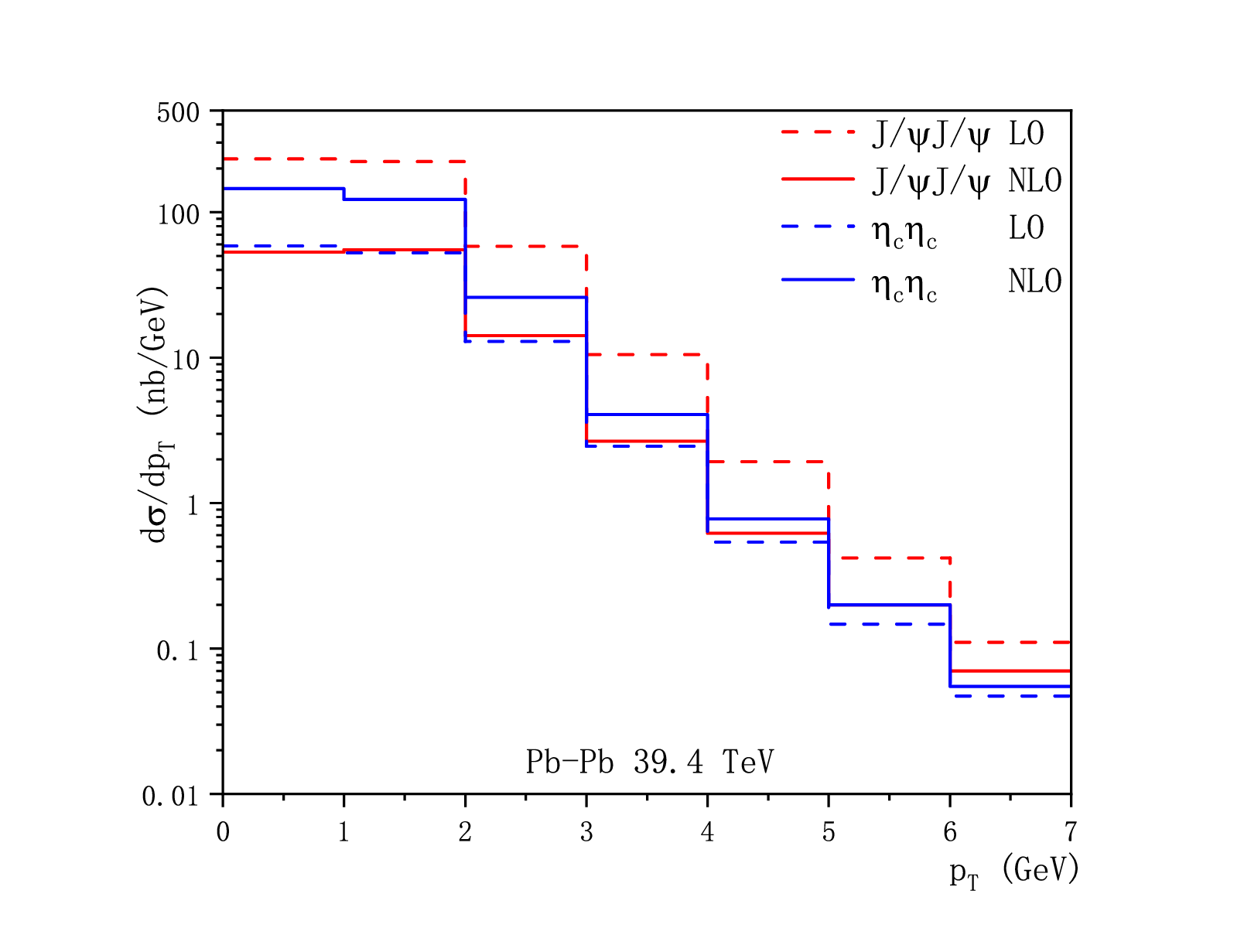}}
	\subfigure{\includegraphics[scale=0.28]{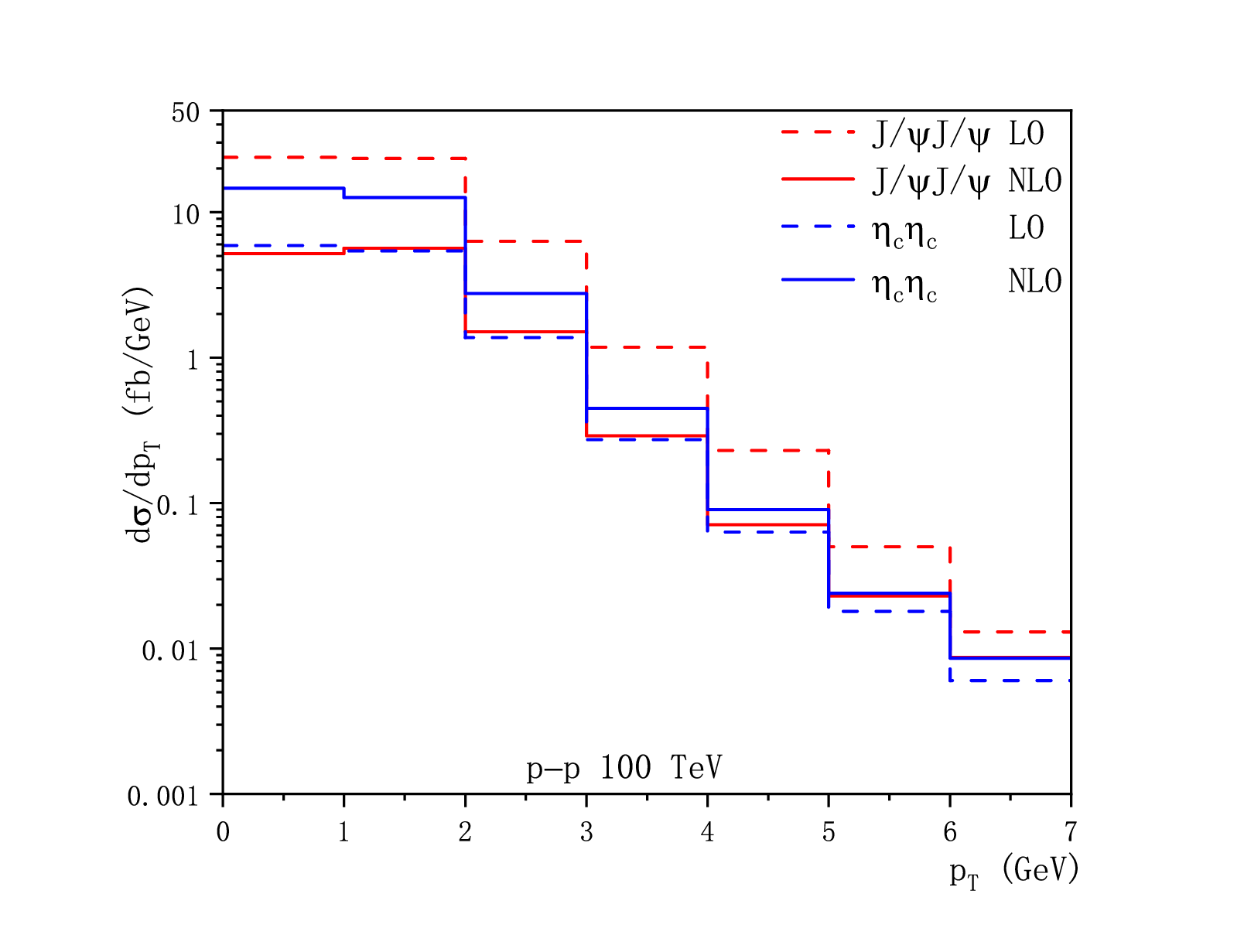}}		
	\caption{The transverse momentum $p_T$ distributions for $J/\psi\mbox{-}J/\psi$ and $\eta_c\mbox{-}\eta_c$ production via ultraperipheral Pb-Pb and p-p collisions.}
	\label{Figpt}
\end{figure}

\begin{figure}[htbp!]			
	\centering
	\subfigure{\includegraphics[scale=0.28]{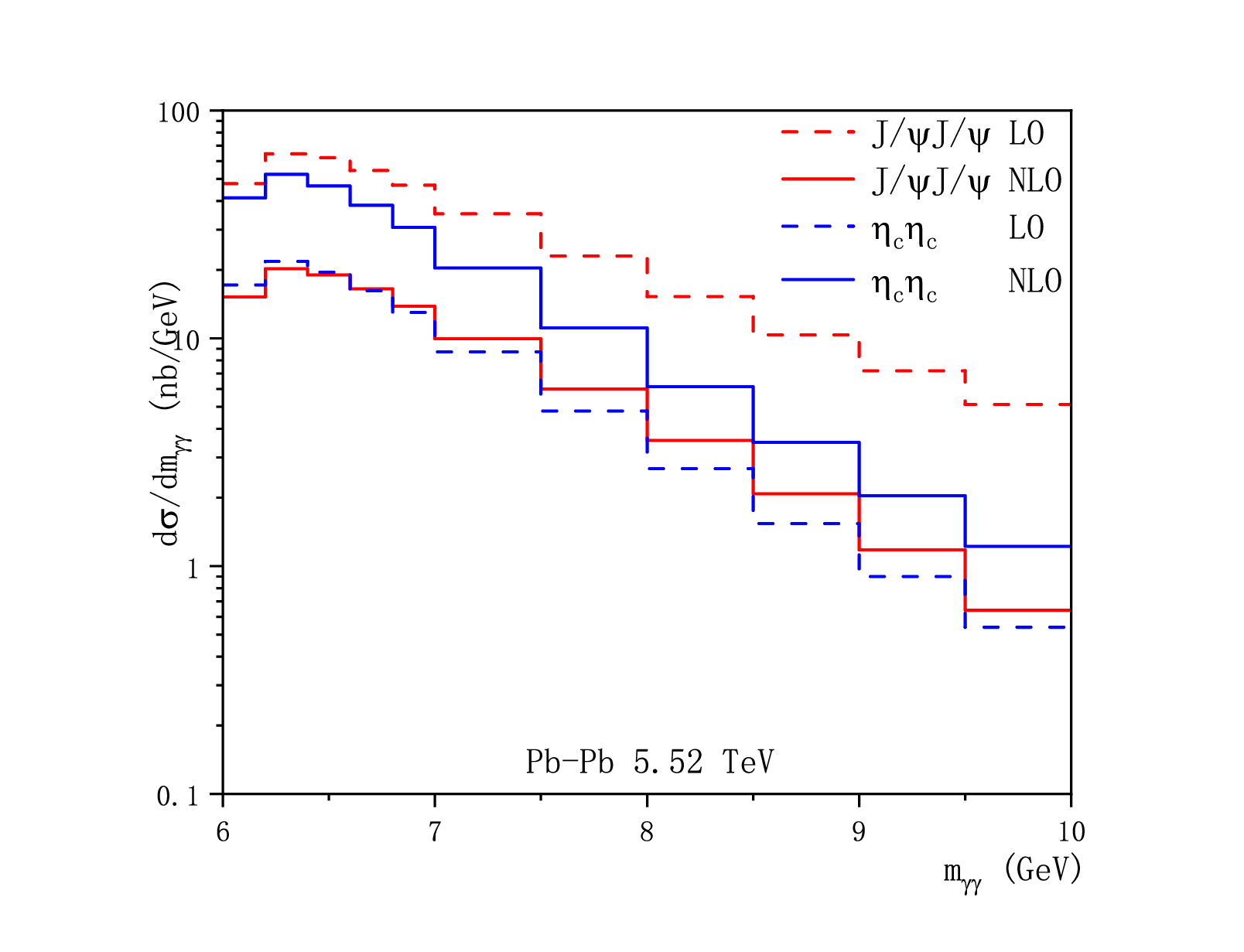}}
	\subfigure{\includegraphics[scale=0.28]{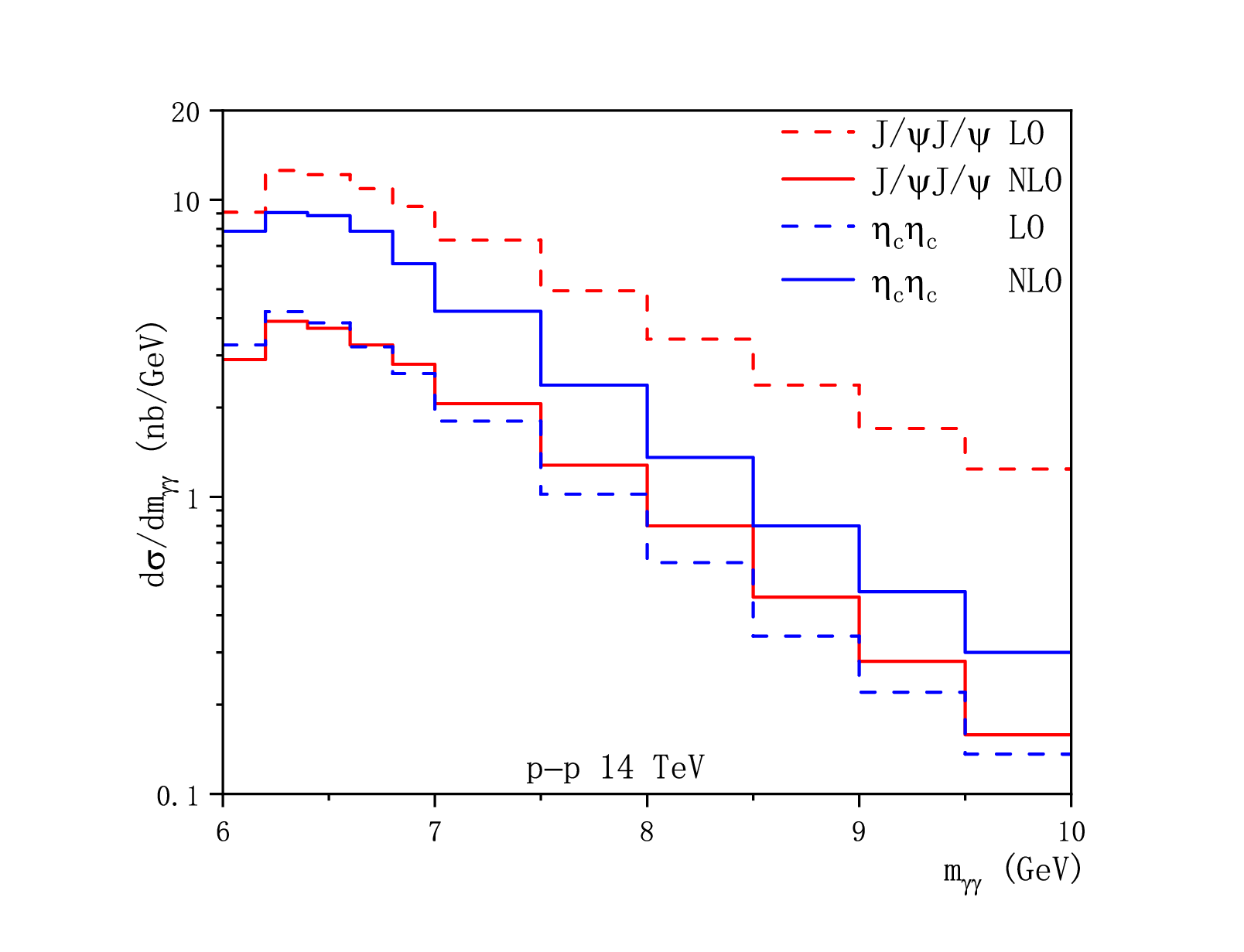}}
	\subfigure{\includegraphics[scale=0.28]{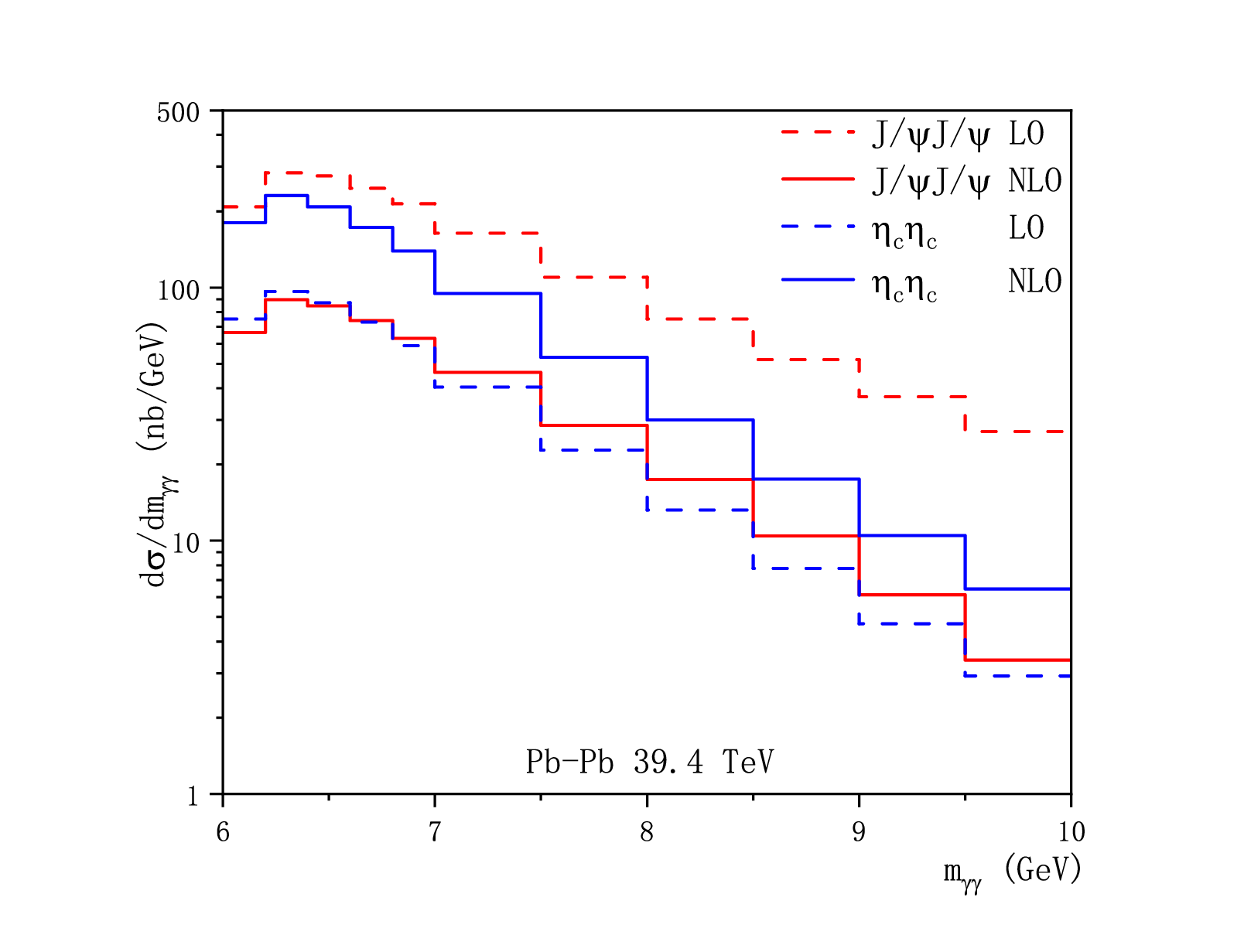}}
	\subfigure{\includegraphics[scale=0.28]{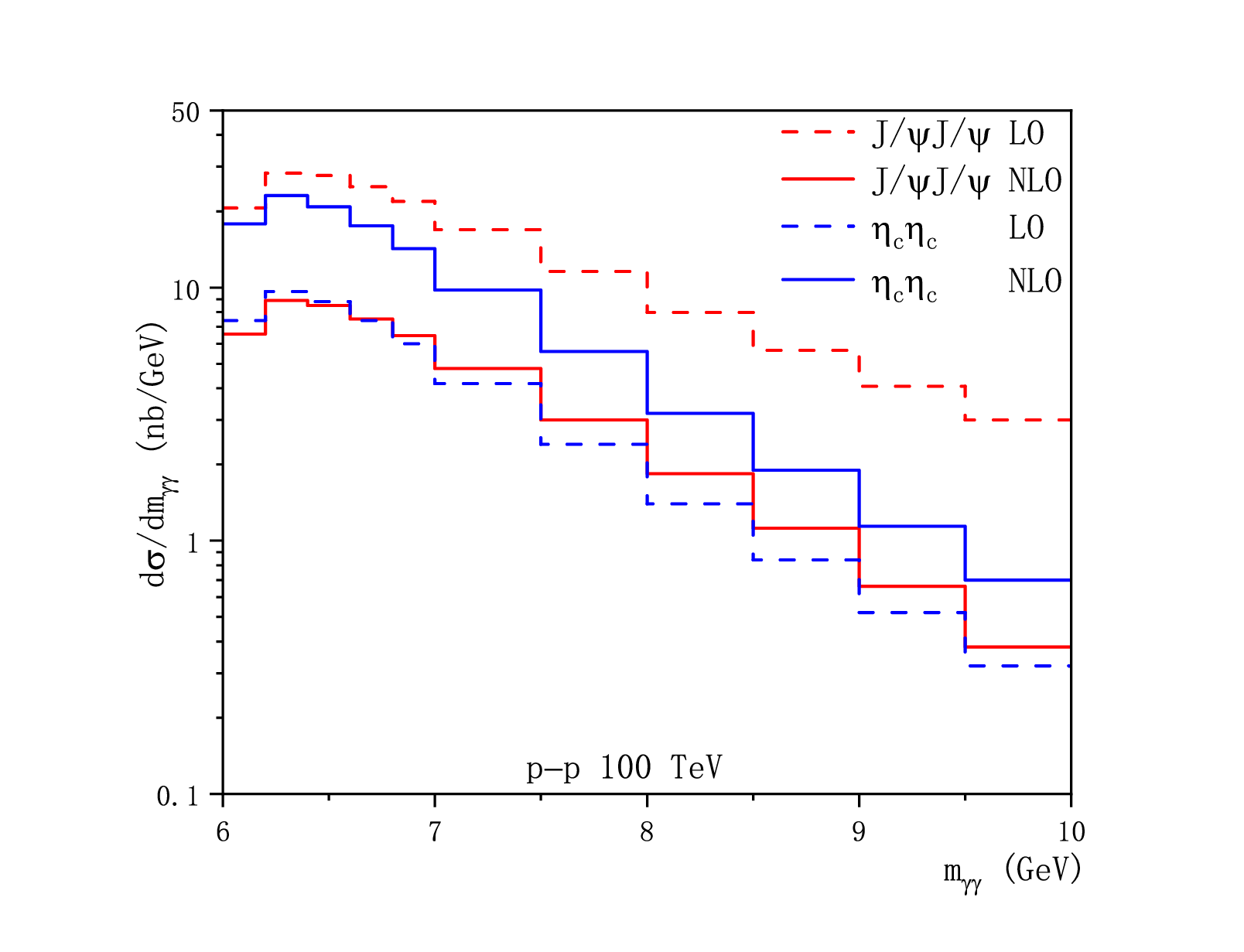}}		
	\caption{The invariant mass $m_{\gamma\gamma}$, i.e., $m_{J/\psi J/\psi}$, distributions for $J/\psi\mbox{-}J/\psi$ and $\eta_c\mbox{-}\eta_c$ production via ultraperipheral Pb-Pb and p-p collisions.}
	\label{Figmass}
\end{figure}

\begin{figure}[htbp!]			
	\centering
	\subfigure{\includegraphics[scale=0.28]{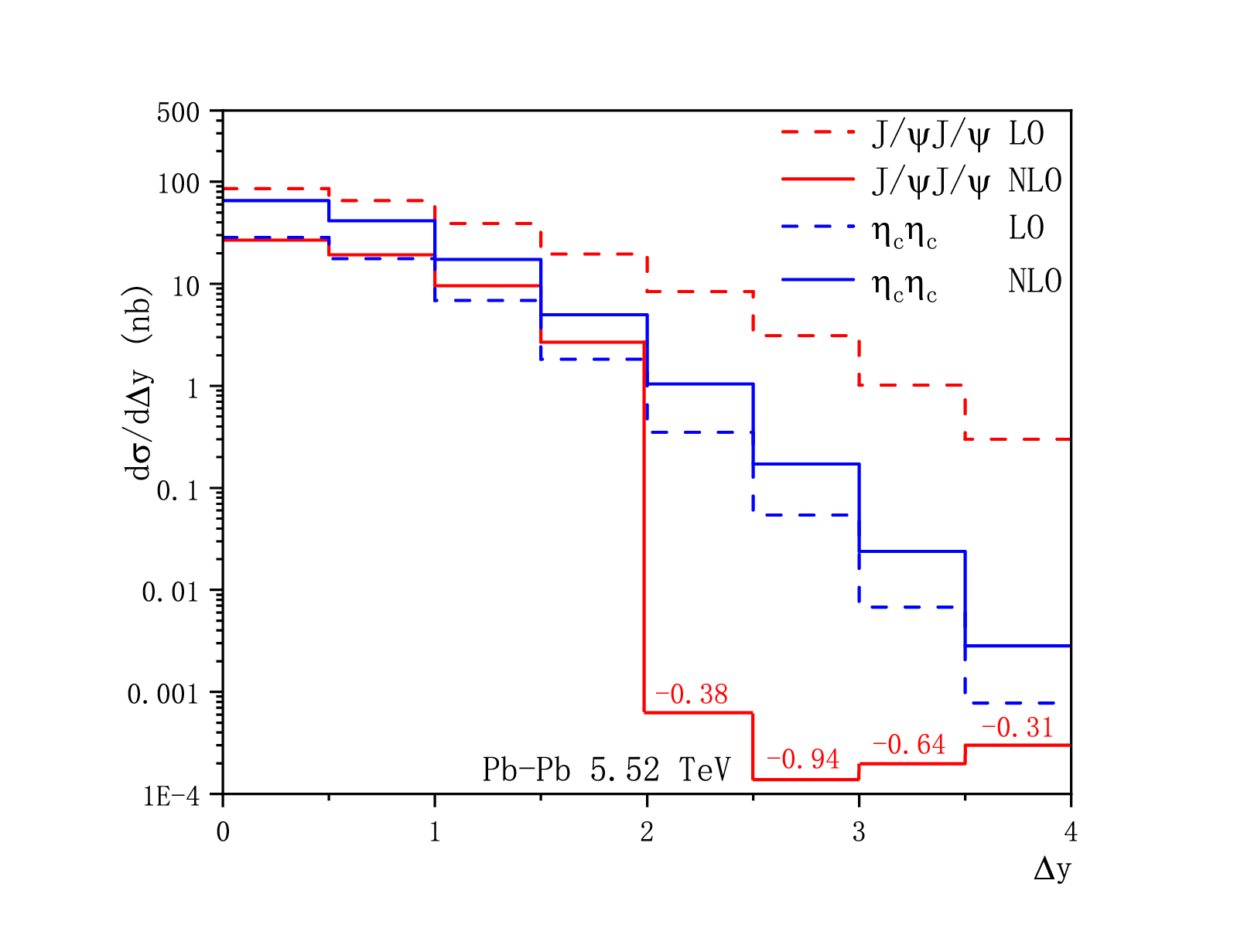}}
	\subfigure{\includegraphics[scale=0.28]{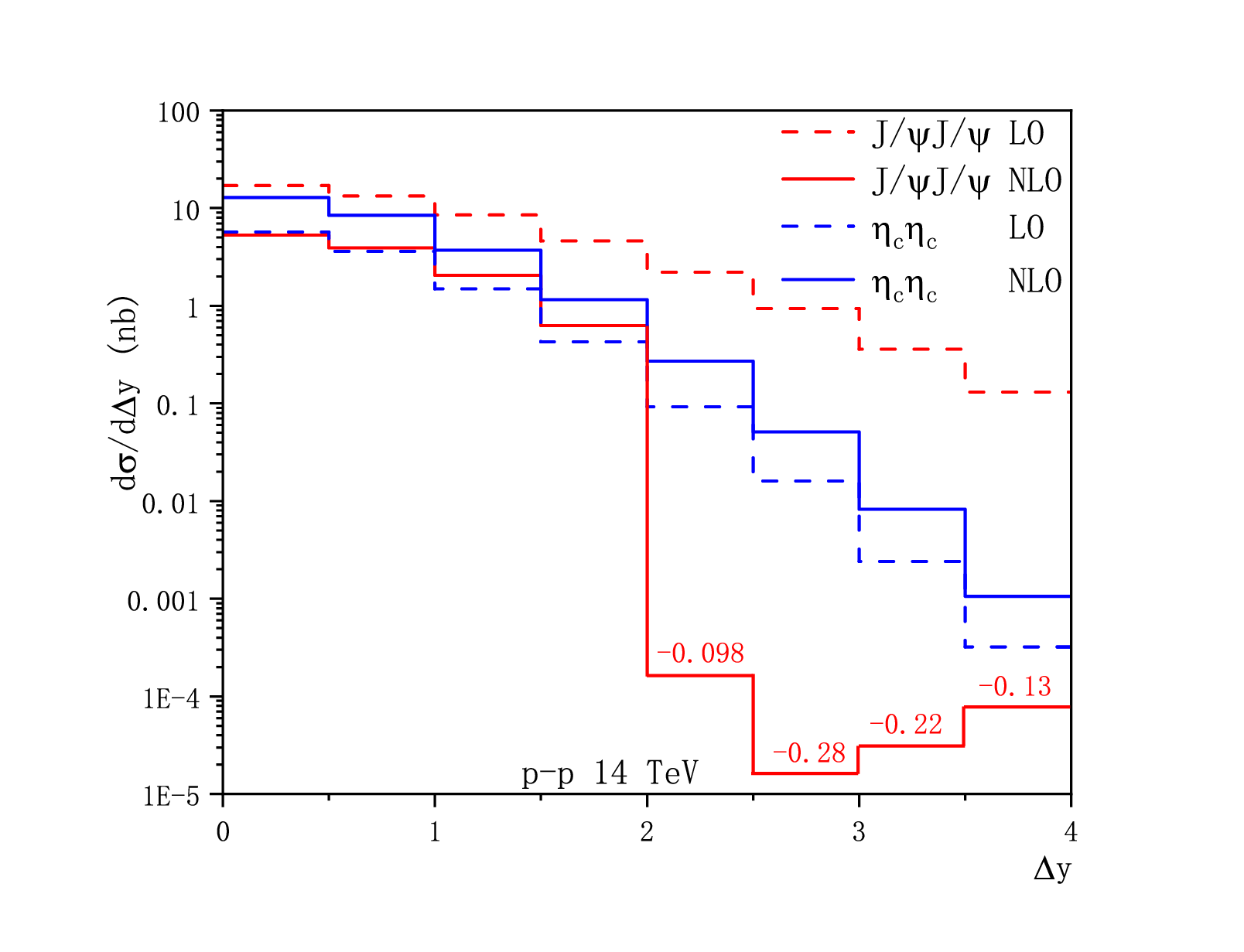}}
	\subfigure{\includegraphics[scale=0.28]{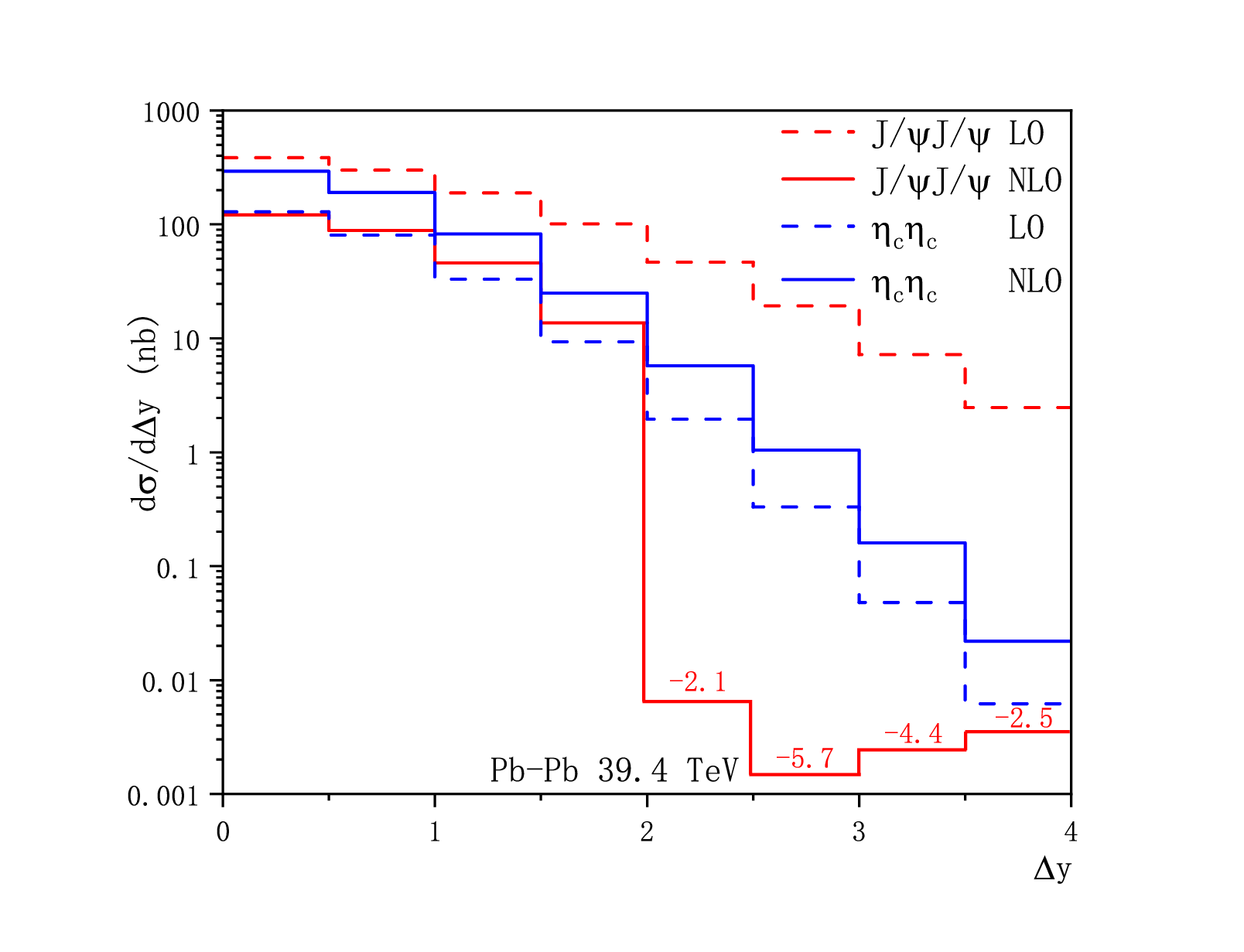}}
	\subfigure{\includegraphics[scale=0.28]{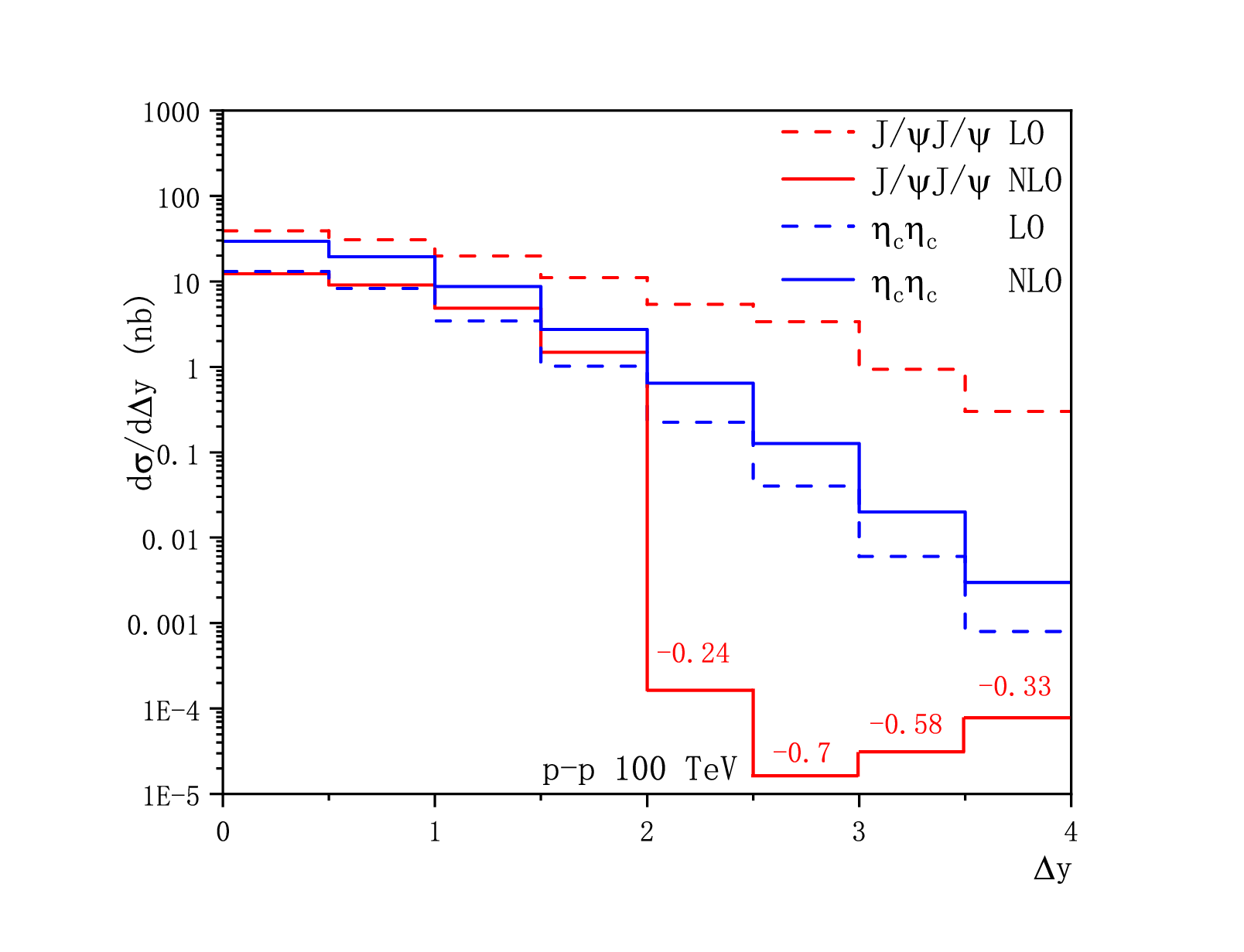}}		
	\caption{The absolute value of rapidity difference of the $J/\psi$-pair,  i.e., $\Delta y=|y^1_{J/\psi}-y^2_{J/\psi}|$, distributions for $J/\psi\mbox{-}J/\psi$ and $\eta_c\mbox{-}\eta_c$ production via ultraperipheral Pb-Pb and p-p collisions.}
	\label{Figy}
\end{figure}

As the number of events corresponding to $\gamma+\gamma\to J/\psi\mbox{-}J/\psi\ (\eta_c\mbox{-}\eta_c)$ are considerable, it is worthy to perform a more elaborate phenomenological analysis. The differential transverse momentum distribution $d\sigma/dp_T$ and invariant mass of charmonium pair $d\sigma/dm_{\gamma\gamma}$ for $J/\psi$-pair and $\eta_c$-pair productions at the HL-LHC and FCC are shown in Fig. \ref{Figpt}, \ref{Figmass} respectively. As the effective photon-photon luminosities decrease dramatically with $W_{\gamma\gamma}$, the main contribution can be only related to small $W_{\gamma\gamma}$ values, therefore we restrict the $p_T$ of charmonium to be 0-7 GeV in the transverse momentum distribution, and the $m_{\gamma\gamma}$ of the charmonium pair is set to be 6-10 GeV. The cross sections decease rapidly versus high $p_T$, showing a logarithmic dependence of $p_T$. The invariant mass distribution reaches the maximum around threshold energy. In FIG.\ref{Figy}, we show the rapidity difference $\Delta y$, the absolute value of the difference of charmonium-pair, distributions via Pb-Pb and p-p ultraperipheral collisions. The produced charmonium-pair mainly shares similar rapidity and decreases rapidly with the rapidity difference. While for the region $\Delta y > 2$, the $J/\psi-J/\psi$ cross sections at NLO are negative, this may be explained by a large negative loop corrections. The unphysical behavior of the NLO charmonium production cross section in UPC could be cured via resetting the renormalization scale, see TABLE \ref{TabScaleVs}. We note that similar negative differential cross section (small $p_T$ or large rapidity) can be found in Refs.\cite{Lansberg:2010cn,Feng:2015cba,ColpaniSerri:2021bla,Chen:2024dkx}, resetting the scale or resummation the logarithms may be useful to alleviate the behavior. A high order correction could also be necessary which deserves further investigation. 
\begin{table}[ht]
	\caption{The NLO (LO) $|\Delta y|$, absolute value of rapidity difference of $J/\psi$-pair, distribution (unit = nb) versus different renormalization scale via Pb-Pb ultraperipheral collision with $\sqrt{s_{NN}}$ = 5.52 TeV.}
	{\footnotesize
	\begin{center}
		\centering
		\begin{tabular}{|m{2.6cm}<{\centering}|m{1.6cm}<{\centering}|m{1.6cm}<{\centering}|m{1.6cm}<{\centering}|m{1.6cm}<{\centering}|m{1.7cm}<{\centering}|m{1.7cm}<{\centering}|m{1.7cm}<{\centering}|m{1.7cm}<{\centering}|}
			\toprule
			\hline
			$|\Delta y|$ & 0-0.5 & 0.5-1 & 1-1.5 & 1.5-2 & 2-2.5 & 2.5-3 & 3-3.5 & 3.5-4\\
			\hline
			$\mu = \sqrt{4m_c^2+p_T^2}$ & 13.4(43.0) & 9.65(32.6) & 4.80(19.6) & 1.34(9.84) & -0.19(4.20) & -0.47(1.55) & -0.32(0.51) & -0.15(0.15) \\
			\hline
			$\mu = \sqrt{\hat{s}}$      & 32.6(25.7) & 24.9(18.8) & 14.9(10.6) & 7.17(4.83) & 2.85(1.85)  & 0.96(0.61)  & 0.28(0.18)  & 0.07(0.047) \\
			\hline		
		\end{tabular}
	\end{center}
    }    
	\label{TabScaleVs}
\end{table}

The fully charm tetraquark state X(6900), which is recently observed by LHCb Collaboration \cite{LHCb:2020bwg} (another two similar resonances around 6552 and 7287 MeV are established by CMS Collaboration \cite{CMS:2023owd}), can be also produced via photon-photon fusion in the UPC and then decays into double charmonium, hence contributes in the phenomenological analysis of double charmonium final state. Using the Low formula \cite{Low:1960wv,Budnev:1975poe,Shao:2022cly}, the production cross section of X(6900) via UPC can be written in terms of the two-photon decay width $\Gamma_{X(6900)\to\gamma\gamma}$ and the effective luminosity $\frac{d\mathcal{L}_{\gamma\gamma}^{AB}}{dW_{\gamma\gamma}}$ at the resonance mass,
\begin{align}
	\sigma(AB\xrightarrow{\gamma\gamma}AB+X(6900)) = 4\pi^2(2J+1)\frac{\Gamma_{X\to\gamma\gamma}}{m^2_X}\left.\frac{d\mathcal{L}_{\gamma\gamma}^{AB}}{dW_{\gamma\gamma}}\right|_{W_{\gamma\gamma}=m_X},
\end{align}
where J is the spin of X(6900), we take J = 0 in the following discussion. The effective luminosity is calculated by
\begin{align}
	\frac{d\mathcal{L}_{\gamma\gamma}^{AB}}{dW_{\gamma\gamma}} = \frac{2}{W_{\gamma\gamma}}\int_{\frac{W^2_{\gamma\gamma}}{s_{NN}}}^{1}\frac{dx}{x}n_{\gamma/A}(x)n_{\gamma/B}(\frac{W^2_{\gamma\gamma}}{xs_{NN}}).
\end{align}
\begin{table}[ht]
	\caption{The estimations of two-photon decay widths and production cross sections in Pb-Pb UPC with c.m. $\sqrt{s_{NN}} = 5.52\ \rm TeV$ corresponding to vector-meson dominance (VMD) \cite{Biloshytskyi:2022dmo,Fariello:2023uvh}, non-relativistic approximation (NRA) \cite{Biloshytskyi:2022pdl} and $\chi_{c0}\to\gamma\gamma$ approximation \cite{Goncalves:2021ytq}. }
	\begin{center}
		\centering
		\begin{tabular}{|m{5.6cm}<{\centering}|m{2.8cm}<{\centering}|m{2.8cm}<{\centering}|m{2.8cm}<{\centering}|}
			\toprule
			\hline
			& VMD & NRA & $\chi_{c0}\to\gamma\gamma$ \\
			\hline
			$\Gamma_{X(6900)\to\gamma\gamma}$ [keV] & 67 & 10 & 2 \\
			\hline
			$\sigma$(PbPb$\to$ PbPb+X(6900)) [nb] & $2\times10^5$ & $3\times10^4$ & $6\times10^3$\\
			\hline 		
		\end{tabular}
	\end{center}    
	\label{Tab6900}
\end{table}

For Pb-Pb collision with nucleon-nucleon c.m. energy $\sqrt{s_{NN}} = 5.52 \ \rm TeV$, the effective two photon luminosity $\left.\frac{d\mathcal{L}_{\gamma\gamma}^{PbPb}}{dW_{\gamma\gamma}}\right|_{W_{\gamma\gamma}=6.9\ \rm GeV} = 9.303\times10^{3} \ \rm GeV^{-1}$, the estimated cross sections are given in TABLE \ref{Tab6900}. Supposing that X(6900) mainly decays into di-charmonium, the produced di-charmonium is much great than the direct $J/\psi J/\psi + \eta_c\eta_c$ yields (93.1 nb). This abnormal behavior may be related to the overestimation of di-photon and di-charmonium decay fractions, which deserves further investigation.

\section{SUMMARY AND CONCLUSIONS}
In this work, we investigate the double charmonium production via photon-photon fusion with various ultraperipheral ion-ion collisions in the framework of NRQCD factorization formalism at NLO accuracy. Both $J/\psi$ and $\eta_c$ are considered in color-singlet configuration. The total cross sections and differential distributions versus transverse momentum, invariant mass and rapidity difference at the HL-LHC and FCC are given. The theoretical uncertainties caused by charm quark mass and renormalization scale are also estimated.

Numerical results show that the NLO QCD corrections are large but negative for $J/\psi$ pair, while positive for $\eta_c$ pair. Based on designed luminosities of each ion-ion collision at the HL-LHC and FCC, a considerable number of double charmonium events can be expected. Due to the event topologies for ultraperipheral collision are vary clear, the background from various QCD interactions can be suppressed, hence the detailed experimental investigation are feasible. The production rate of exotic fully-charmed tetraquark X(6900) through Pb-Pb UPC is also discussed via Low formula with different diphoton widths approximation. 

\vspace{1.4cm} {\bf Acknowledgments}
We thank Long-Bin Chen for useful discussion. This work is supported by the National Natural Science Foundation of China (NSFC) under the Grants
Nos. 12205061, 12275185 and 12335002.

\end{document}